\documentclass{PoS}

\usepackage[english]{babel}
\usepackage[utf8]{inputenc}
\usepackage[T1]{fontenc}
\usepackage{amsmath,amssymb,braket,slashed,mathrsfs}

\usepackage[squaren,Gray,cdot,binary]{SIunits}
\usepackage{graphicx}
\graphicspath{{./figures/}}
\usepackage[numbers,square]{natbib}

\newcommand{\cf}{\textit{cf.}~}
\newcommand{\eg}{\textit{e.g.}~}
\newcommand{\ie}{\textit{i.e.}~}
\renewcommand{\bar}{\overline}
\DeclareMathOperator{\bigo}{O}
\DeclareMathOperator{\BZ}{BZ}

\newcommand{\diff}{\mathrm{d}}
\newcommand{\dm}{\delta m}
\newcommand{\DM}{\Delta M}
\newcommand{\DQCDM}{\Delta_{\QCD}M}
\newcommand{\DQEDM}{\Delta_{\QED}M}
\renewcommand{\epsilon}{\varepsilon}
\newcommand{\QCD}{\mathrm{QCD}}
\newcommand{\QED}{\mathrm{QED}}
\newcommand{\QEDL}{\mathrm{QED}_{\mathrm{L}}}
\newcommand{\QEDTL}{\mathrm{QED}_{\mathrm{TL}}}
\DeclareMathOperator{\SU}{SU}

\newcommand{\eV}{\electronvolt}
\newcommand{\fm}{\femto\meter}
\renewcommand{\eqref}[1]{(\ref{#1})}
\newcommand{\secref}[1]{section \ref{#1}}
\newcommand{\tabref}[1]{Table \ref{#1}}
\newcommand{\figref}[1]{Figure \ref{#1}}

\title{Inclusion of isospin breaking effects\\in lattice simulations}
\ShortTitle{Inclusion of isospin breaking effects in lattice simulations}
\author{\speaker{Antonin Portelli}\\
        School of Physics \&\ Astronomy, University of Southampton, 
        SO17 1BJ, UK\\
        E-mail: \email{a.portelli@soton.ac.uk}}
\abstract{Isospin symmetry is explicitly broken in the Standard Model by the
mass and electric charge of the up and down quarks. These effects
represent a perturbation of hadronic amplitudes at the percent level. Although
these contributions are small, they play a crucial role in hadronic and nuclear
physics. Moreover, as lattice computations are becoming increasingly precise,
it is becoming more and more important to include these effects in numerical
simulations. We summarize here how to properly define QCD and QED on a finite
and discrete space-time so that isospin corrections to hadronic observables can
be computed \emph{ab-initio} and we review the main results on the isospin
corrections to the hadron spectrum. We mainly focus on the recent work going
beyond the electro-quenched approximation.}
\FullConference{The 32nd International Symposium on Lattice Field Theory,\\
		23-28 June, 2014\\
		Columbia University New York, NY}

\begin{document}
\section{Motivation}
In an isospin symmetric world, the up ($u$) and down ($d$) quarks are identical
particles. It is known (\cf\tabref{tab:udquarks}) than in Nature isospin
symmetry is explicitly broken by the non-zero mass and electric charge
differences of the $u$ and $d$ quarks. However, the effects of this breaking
are expected to be small relative to typical strong interaction energies such
as hadron masses. Indeed, it is clear that the light quark mass mass difference
$\dm=m_u-m_d$ represents one percent or less of any typical QCD energy scale.
Similarly, the strength of the electromagnetic (EM) interaction relatively to
the strong one is essentially given at low energy by the fine structure
constant $\alpha\simeq 0.007$. For those reasons we can reasonably state that,
for observables with a non-vanishing isospin symmetric part, isospin symmetry
is a good approximation of reality with an $\bigo(1\%)$ relative error.
\begin{table}[b]
    \begin{center}
	\begin{tabular}{|l|c|c|}
		\cline{2-3}
		\multicolumn{1}{c|}{~} & $u$ & $d$                  \\
		\hline
		Mass ($\unit{\!}{\mega\eV}$)~\citep{PDG2014}& 
        $2.3\left(^{+0.7}_{-0.5}\right)$ & 
		$4.8\left(^{+0.7}_{-0.3}\right)$                    \\
		Charge & $\frac{2}{3}e$ & $-\frac{1}{3}e$           \\	
		\hline
	\end{tabular}
    \end{center}
    \caption{Physical properties of the up and down quarks.}
    \label{tab:udquarks}
\end{table}

Nevertheless, it is interesting to note that these small isospin breaking
corrections are crucial to describe the structure of atomic matter in the
Universe. Indeed, one particular effect of isospin symmetry breaking is the
mass splitting between the proton ($p$) and the neutron ($n$). This mass
difference is known experimentally with an impressive accuracy~\citep{PDG2014}:
\begin{equation}
    \DM_N = M_n-M_p = \unit{1.2933322(4)}{\mega\eV}
\end{equation}
The sign of this splitting makes the proton and the hydrogen atom
stable physical states. Also, the size of $\DM_N$ determine the phase space
volume for the neutron $\beta$-decay \mbox{$n\to p+e^-+\bar{\nu}_e$}. At early
times of the Universe ($t\sim\unit{1}{\second}$ and $T\sim\unit{1}{\mega\eV}$)
and under standard assumptions\footnote{The neutrino number density
$n_{\nu}/n_{\gamma}$ is assumed to have the order of the baryon density number
which is very small. This assumption is not valid anymore in some new physics
scenarios but even in these hypothetical cases $n_n/n_p$ depends strongly on
$\DM_N$.}, the existence of $\beta$-decay allows to infer that the ratio of the
number of neutrons and protons is approximatively equal to:
\begin{equation}
    \frac{n_n}{n_p}\simeq\exp\left(\frac{\DM_N}{T}\right)
\end{equation}
This ratio is one important initial conditions of Big Bang Nucleosynthesis.
Also, in our actual Universe, $\beta$-decay and its inverse process are known
to be responsible for the generation of a large majority of the stable nuclides
chart though nuclear transmutation. Even if the nucleon isospin mass splitting
is a well known quantity, predicting it from first principles is a difficult
problem because of the complex non-perturbative interactions of quarks inside
the nucleon. The proton carries an additional EM self-energy compared to the
neutron, so just from QED one would expect to have $\DM_N<0$. However, the fact
that the experimental value of $\DM_N$ has the opposite sign suggests that the
strong isospin breaking effects are competing against the EM effects with a
larger magnitude. This would mean that an important part of the structure of
nuclear matter as we know it relies on a subtle cancellation between the small
EM and strong breaking effects of isospin symmetry in the nucleon system.
Therefore, it is fundamental to have a theoretical understanding of the nucleon
isospin mass splitting.

Considering that isospin breaking effects in the hadron mass spectrum are
generally measured quite precisely, it is also interesting to understand how
one can use this information to deduce the masses of the individual $u$ and $d$
quark masses. For example, it is important to know if $m_u=0$ could be a
realistic solution to the strong CP problem. While recently (\cf the FLAG review
\citep{FLAG2014}) considerable progress has been made in determining
precisely the average up-down quark mass $m_{ud}$ from first principles, such
a computation is still missing for the individual masses. Because the kaon
($K$) is a pseudo-Goldstone boson of chiral symmetry breaking, the isospin mass
splitting $\DM^2_K=M_{K^+}^2-M_{K^0}^2$ is very sensitive to $\dm$. But in
order to extract $\dm$, one has to understand how to subtract the EM
contribution to this splitting. One well known result in this direction is
Dashen's theorem~\citep{Dashen:1969tn} which states that, in the $\SU(3)$ chiral
limit, the EM Kaon splitting is equal to the EM pion ($\pi$) splitting:
\begin{equation}
    \DQEDM^2_K=\DQEDM^2_{\pi}+\bigo(\alpha m_s)
    \label{eq:dashth}
\end{equation}
This result is important because it is known \citep{FLAG2014} that with good
accuracy, $\DQEDM^2_{\pi}\simeq\DM^2_{\pi}$. The remaining question is: how
large are the $\bigo(\alpha m_s)$ corrections in \eqref{eq:dashth}? One way to
quantify these corrections is to consider the dimensionless quantity $\epsilon$
defined in~\citep{FLAG2014} as follows:
\begin{equation}
    \epsilon=\frac{\DQEDM^2_K-\DQEDM^2_{\pi}}{\DM^2_{\pi}}
    \label{eq:eps}
\end{equation}
This quantity is constructed such that it vanishes in the $\SU(3)$ chiral limit.
There were several attempts in the 1990s to compute these corrections
analytically from effective theories which leaded to controversial results.
It makes this quantity a good target for a lattice calculation.

The problems presented in this section, and more generally in any computation
of isospin corrections to low-energy QCD observables, are difficult to solve
because of the highly non-perturbative behavior of the strong interaction in
this regime. It has been shown
\citep{FLAG2014} that it is now possible to predict fundamental
isospin symmetric QCD observables through lattice QCD simulations with a full
control over the method's uncertainties. It is then reasonable to think that
lattice simulations could be a reliable way to understand and compute isospin
breaking effects. Moreover, besides the physical interest of these effects,
actual lattice calculations are reaching a sub-percent precision on several
standard observables and the assumption of isospin symmetry is becoming the
dominant source of systematic uncertainty.
\section{Lattice QCD+QED}
In this section we review how to add EM interactions to lattice simulations. As
we will see, the main difficulties comes from the singular infrared structure
of QED. We explain a possible way to define QED in a finite volume and what are
the associated finite-size effects. We then discuss the discretization of the
theory and the simulation techniques used so far.
\subsection{QCD+QED in a finite volume}
\label{ssec:fv}
Let us consider a diagrammatic
contribution $\mathscr{D}$ to a correlation function featuring a photon loop
(\eg the 1-loop part of the electron EM self-energy). In infinite volume,
$\mathscr{D}$ will have the following form:
\begin{equation}
    \mathscr{D}_{\infty}=
    \int\frac{\diff^4 k}{(2\pi)^4}\frac{1}{k^2}f(k,p_1,\dots,p_n)
    \label{eq:irdiv}
\end{equation}
The integral \eqref{eq:irdiv} may be ultraviolet (UV) divergent, which can be
dealt with through renormalization. The photon pole at $k^2=0$ can also generate
infrared (IR) divergences. However, \eqref{eq:irdiv} is not mathematically
undefined \textit{per se}, the undefined $k^2=0$ value of the integrand is just
a point (set of measure $0$) and can be ignored. Moreover, it is known that in
some cases (\eg the on-shell self-energy of a particle), this singularity is
integrable. In a finite volume with temporal extent $T$ and spatial extent $L$,
momenta become quantized in a way depending on the choice of boundary
conditions. If one chooses periodic boundary conditions for the photon field,
then the contribution \eqref{eq:irdiv} becomes:
\begin{equation}
  \mathscr{D}(T,L)=
  \frac{1}{TL^3}\sum_{k\in\BZ(T,L)}\frac{1}{k^2}f(k,p_1,\dots,p_n)
  \label{eq:irdivfv}
\end{equation}
where:
\begin{equation}
  \BZ(T,L)=\frac{2\pi}{T}\mathbb{Z}\times\frac{2\pi}{L}\mathbb{Z}^3
\end{equation}
Now the expression \eqref{eq:irdivfv} has an isolated, undefined contribution
coming from the photon pole which cannot be summed in any way. As mentioned in
\citep{Portelli:2010tz,Davoudi:2014kp}, this singularity is classically related
to the fact that Gauss' law does not authorize a net charge to exist in a
finite, periodic volume.

\subsubsection{Photon zero-mode subtraction schemes}
If one wants to keep periodic boundary conditions on the photon field, one
possible solution to deal with the zero-mode singularity is to remove a subset
of mode containing $0$ from the finite-volume degrees of freedom. This will of
course alter the physics in finite volume, but if the chosen subset converges
in the infinite volume limit to a set of measure $0$ then naively the physics
in infinite volume remains unchanged. This is only a necessary condition, in
principle one needs to check that the subtraction procedure does not
accidentally couple the IR and UV structure of the theory which could introduce
a complicated volume-dependent renormalization of the theory.

Naively, the most minimal zero-mode subtraction procedure is to set the $k=0$
mode of the photon field to $0$. Following~\citep{Borsanyi:2014jba}, we name
the resulting theory $\QEDTL$. This finite-volume prescription has been used
for numerical calculations in
\citep{Duncan:1996cy,deDivitiis:2013er,Borsanyi:2013va}. Although this scheme
is simple, it introduces some strong finite-volume effects which can be hard to
control. Indeed, considering the $T\to+\infty$ limit at fixed $L$ of the
$\QEDTL$ version of \eqref{eq:irdivfv}, one obtains:
\begin{equation}
  \mathscr{D}_{\QEDTL}(T,L)=\frac{1}{TL^3}
  \sum_{\substack{k\in\BZ(T,L)\\k\neq 0}}f(k,p_1,\dots,p_n)
  \underset{T\to+\infty}{\longrightarrow}
  \frac{1}{L^3}\int\frac{\diff k_0}{2\pi}\sum_{\mathbf{k}\in\BZ(L)}
  f(k,p_1,\dots,p_n)
  \label{eq:qedtllarget}
\end{equation}
where $\BZ(L)=\frac{2\pi}{L}\mathbb{Z}^3$. Because it is one-dimensional, the
integral in \eqref{eq:qedtllarget} might be IR divergent even in cases where
its four-dimensional version converges. As an example, finite-volume effects on
the 1-loop mass correction in spinor $\QEDTL$ were computed in
\citep{Borsanyi:2014jba}:
\begin{align}
m_{\QEDTL}(T,L)\underset{T,L\to+\infty}{=} 
m&\left\{1-q^2\alpha\left[\frac{\kappa}{2mL}
\left(1+\frac{2}{mL}\left[1-\frac{\pi}{2\kappa}{\frac{T}{L}}\right]\right)
\right.\right.\notag\\
&
\quad\left.\left.-\frac{3\pi}{(mL)^3}
\left[1-\frac{\coth(mT)}{2}\right]-
\frac{3\pi}{2(mL)^4}\frac{L}{T}\right]\right\}
\label{eq:mqedtl}
\end{align}
where $m$ is the infinite volume mass, $q$ is the charge in units of $e$ and
$\kappa=2.83729\dots$ is a known numerical constant. This expansion is exact up
to corrections that decay exponentially in the infinite volume limit. In this
example we explicitly see a term proportional to $\frac{T}{L^3}$ which
represents the IR divergence related to the limit \eqref{eq:qedtllarget}. In
conclusion, it appears that $\QEDTL$ has two cumbersome properties. Firstly,
the infinite-volume limit has to be taken with special care (\ie by keeping
$\frac{T}{L^3}$ bounded) and secondly this extrapolation depends on the aspect
ratio $\frac{T}{L}$. As discussed in~\citep{Borsanyi:2014jba}, the singularity
of $\QEDTL$ in the large $T$ limit can be explained in the following way. The
photon zero-mode removal can be implement by adding a non-local term in the
Lagrangian of the theory. This term couples values of the electromagnetic
potential on different time-slices, breaking the reflection positivity of the
action. So strictly speaking $\QEDTL$ does not admit a quantum mechanical
description and the divergence in $T$ is a symptom of the lack of a
thermodynamic limit. This singular behavior has been discovered independently
by the MILC collaboration \citep{Basak:2014td}.

An alternative to $\QEDTL$ is to remove all spatial zero-modes, \ie to set
to zero all the photon modes $k$ with $\mathbf{k}=0$. This scheme is inspired
from~\citep{Hayakawa:2008ci} where QED is formulated in a finite spatial volume
directly with an infinite temporal dimension. We denote this prescription
$\QEDL$~\citep{Borsanyi:2014jba}. Because it does not couple field values on
different time-slices, $\QEDL$ has positive reflexivity and therefore a correct
particle physics interpretation. In this theory, the finite-volume effects on
the 1-loop mass correction of a spin $\frac12$ particle are given by
\citep{Borsanyi:2014jba}:
\begin{equation}
  m_{\QEDL}(T,L)\underset{T,L\to+\infty}{=} 
  m\left\{1-q^2\alpha
  \left[\frac{\kappa}{2mL}\left(1+\frac{2}{mL}\right)
  -\frac{3\pi}{(mL)^3}\right]\right\}
  \label{eq:mqedl}
\end{equation}
Compared to \eqref{eq:mqedtl}, this relation is now completely independent from
the aspect ratio $\frac{T}{L}$. $\QEDL$ has been used in electro-quenched
simulations in~\citep{Blum:2010ge} and in full simulations in
\citep{Borsanyi:2014jba}.

\begin{figure}[t]
  \centering
    \includegraphics[scale=0.6]{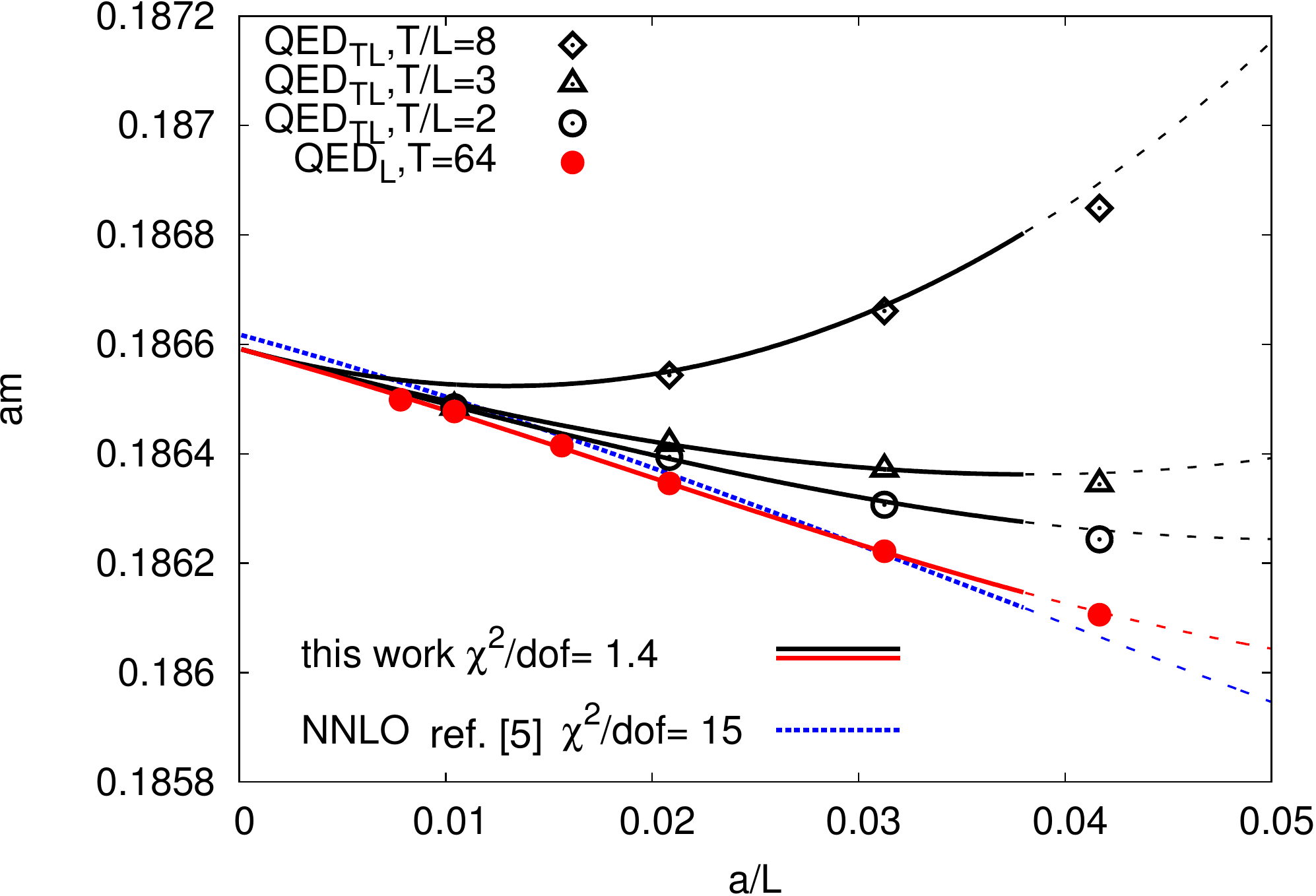}
  \caption{Finite-volume corrections to an elementary fermion mass in QED. The
  black (resp. red) points represent $\QEDTL$ (resp. $\QEDL$) quenched QED
  simulations (which is exact at $\bigo(\alpha)$). The black (resp. red) curves
  represent the theoretical prediction \eqref{eq:mqedtl} (resp.
  \eqref{eq:mqedl}). The only fit parameter is the infinite-volume mass. The
  dashed blue line is the prediction from~\citep{Davoudi:2014kp}, the
  disagreement between this formula and the data is commented in
  \secref{sssec:hadronfv}.}
  \label{fig:qedfvol}
\end{figure}

The mass corrections in $\QEDL$ and $\QEDTL$ were compared to numerical
quenched QED (which is exact at the 1-loop) simulations in
\citep{Borsanyi:2014jba} and perfect agreement is found between the
simulations, \eqref{eq:mqedtl} and \eqref{eq:mqedl}. These results are
summarized in \figref{fig:qedfvol}.

\subsubsection{Finite-volume effects on hadron masses}
\label{sssec:hadronfv}
All the previous statements were made for elementary particles which interact
only through QED. Here we review how to generalize this discussion for hadrons,
\ie composite bound states of the strong interaction. All of the work presented
here are computations performed in $\QEDL$ with an infinite time dimension.

To study finite-volume effects on hadrons masses, one possible approach is to
use low-energy effective theories of the strong interaction coupled to QED.
This was done first for meson masses in the context of $\SU(3)$
partially-quenched chiral perturbation theory in~\citep{Hayakawa:2008ci}. The
results of that work were studied numerically and generalized to an $\SU(2)$
plus heavy kaons theory in~\citep{Blum:2010ge}. EM finite-volume corrections to
meson, baryon, nuclei masses and to the hadronic vacuum polarization were
studied in~\citep{Davoudi:2014kp} using non-relativistic effective theories. In
this context, the finite-volume corrections appear as the elementary particle
ones plus terms depending on the structure of the particle (radius,
polarizabilities, \dots).

However, there is a disagreement between the point-like limit
of~\citep{Davoudi:2014kp} and the QED prediction \eqref{eq:mqedl} concerning
the fermion mass correction. The difference is a relative factor of $2$ in the
$\bigo[\alpha/(mL)^3]$ correction. The QED simulations presented in
\figref{fig:qedfvol} strongly favor \eqref{eq:mqedl}. An explanation for this
discrepancy has been recently proposed in~\citep{Fodor:2015wk}. In the
non-relativistic limit, the particle and antiparticle degrees of freedom
decouple and therefore in non-relativistic theories one does not expect the
antiparticle modes to contribute to the particle mass corrections. However, as
pointed out in~\citep{Fodor:2015wk}, these modes contribute through the
subtraction of the photon zero-modes in $\QEDL$. Once properly added, this
residual fermion-antifermion interaction generates a $\bigo[\alpha/(mL)^3]$
finite-volume correction which solves exactly the discrepancy with
\eqref{eq:mqedl}.

The structure of finite-volume corrections to hadron masses was also studied
beyond the effective theory level~\citep{Borsanyi:2014jba}. In that work it is
confirmed that the two first orders of the finite-volume expansion in
$\frac{1}{mL}$ are universal and identical to the pure QED case. These terms
are determined by gauge invariance, through the constraints on the
electromagnetic form factors provided by the Ward-Takahashi identities, and
follow from the analyticity properties of 1-particle irreducible Green's
functions in the relevant quantum field theories. The structure contributions
only enter at least at order $\bigo[\alpha/(mL)^3]$ which can be seen
explicitly for the effective theories presented
in~\citep{Hayakawa:2008ci,Blum:2010ge,Davoudi:2014kp}. The universality of the
two first orders is an important information as it allows to impose these
corrections analytically in lattice data analyses without introducing any model
dependence.
\subsection{Lattice QED}
There are essentially two approaches to discretize QED: a naive, non-compact
action where the gauge potential $A_{\mu}$ is still the field variable or a
compact Wilson action similar to lattice QCD. By naive discretization we mean
that the lattice action is defined as follows:
\begin{equation}
  S[A_{\mu}]=\frac{a^4}{4}\sum_{x,\mu,\nu}
  [\partial_{\mu}A_{\nu}(x)-\partial_{\nu}A_{\mu}(x)]^2
\end{equation}
where $a$ is the lattice spacing and $\partial_{\mu}$ is some first-order
finite-difference operator. So far, except the starting project of the MILC
collaboration~\citep{Zhou:2014tc}, only the non-compact action has been used in
the context of lattice QCD+QED simulations. One of the main motivations in
making this choice comes from the fact that the non-compact action is free
and still gauge-invariant\footnote{this is not the case for non-compact lattice
QCD~\citep{Seiler:1984ju}}. On the other hand, the compact QED action
introduces photon-photon interactions which are a pure discretization effect.
However, there is in principle no conceptual problem in using compact QED. In
both cases (although it is mandatory for the non-compact action) gauge fixing
needs to be considered. Indeed, with electromagnetic interactions one might be
interested in gauge variant amplitudes involving charged particles. In the
non-compact case, gauge fixing is straightforward for Coulomb and Feynman
gauges~\citep{Borsanyi:2014jba}. Moreover, as explained
in~\citep{Borsanyi:2014jba}, the non-compact action offers an interesting
opportunity for Fourier acceleration of the hybrid Monte-Carlo (HMC) algorithm.
The argument goes as follows: because the pure gauge theory is free, one can
find a distribution for the HMC momenta which exactly cancels any
autocorrelation in the Markov chain. Of course this is not correct anymore once
quarks and gluons are coupled to the system. However, because of the weak
coupling of QED it has been observed that using this particular momentum
distribution still considerably reduces the autocorrelations coming from EM
interactions.
\subsection{QCD+QED simulations}
Up to now, essentially two approaches have been used to perform lattice QCD+QED
simulations. On the one hand, one can use the so-called electro-quenched
approximation which consists in neglecting the EM contribution to the fermionic
determinant (\ie the sea quarks are electrically neutral). This approach is
more cost effective but it is not possible to control reliably the quenching
effects. On the other hand, one can consider the full theory. In the past
couples of year several groups worked or started working on such simulations. We
summarize below the effort done in both approaches.
\subsubsection{Electro-quenched approximation}
From the Monte-Carlo simulation point of view, the coupling of quenched QED
(qQED) to QCD is fairly straightforward. For a given lattice QCD gauge
configuration, one generates a pure gauge QED configuration (which is simply
Gaussian distributed for the non-compact lattice action). Then the QED field is
used to phase the QCD gauge links in the lattice Dirac operator inverted to
obtain the valence quark propagators. Another approach has been used in
\citep{deDivitiis:2013er}: in that work the leading QED corrections are
expressed as pure QCD expectation values in a perturbation theory fashion. In
that framework, the electro-quenched approximation consist in neglecting the
disconnected quark diagrams where the EM currents are self-contracted.

It is easy to show that the missing contributions to the fermionic determinant
are suppressed by both the number of colors and $\SU(3)$ flavor symmetry. Using
this fact and naive dimensional analysis, a quantity computed in the
electro-quenched approximation is expected to suffer from a $\bigo(10\%)$
quenching effect relatively to its electromagnetic
corrections~\citep{Borsanyi:2013va}. Also, partially quenched chiral
perturbation theory allows to provide estimations consistent with the
dimensional one for the light meson mass splittings
\citep{Bijnens:2007jk,Portelli:2012vz}.

A summary of the different lattice QCD+qQED simulation projects can be found in
\citep{Portelli:2013wf}. Apart from a slight update from the MILC group
\citep{Basak:2014td}, the actual situation is essentially identical concerning
electro-quenched simulations.
\subsubsection{Full QCD+QED simulations}
There are essentially three possible ways to compute full QCD+QED correlation
functions. Firstly, it is possible to compute directly the ratio of the QCD+QED
to QCD fermionic determinant for each QCD configuration. These ratios can be
then used to re-weight electro-quenched data. This technique was first proposed
in~\citep{Duncan:2005cq} and applied in exploratory calculations
\citep{Ishikawa:2012iw,Aoki:2012uu}. We see one major limitation of
re-weighting techniques applied to QCD+QED: as the volume gets larger, the
computational cost of the weights increases rapidly and the signal over noise
ratio decreases. As discussed in \secref{ssec:fv}, it is important to
reach large physical volumes in order to control the large finite-volume
effects generated by QED. 

A second approach already mentioned in the previous section is to perform a
perturbative expansion in $\alpha$ and express the QED correction as pure QCD
observables. The determinant contribution then appears as disconnected quark
diagrams. The computation of these diagrams was never attempted in the context
of computing EM corrections to hadronic amplitudes. However, identical diagrams
contribute in other problems and they appear to be extremely difficult to
compute. One can for example look at the recent study by the Mainz group
\citep{Francis:2014th} of the disconnected contributions to the hadronic vacuum
polarization. This approach has the advantage of isolating specific
perturbative contributions which can be useful for the control and
understanding of IR divergences in hadronic processes~\citep{Carrasco:2015uo}.

Finally, one can generate new field configurations including both QCD and QED
actions in the HMC process. Both QCDSF~\citep{Horsley:2013ti, Schierholz} and
MILC~\citep{Zhou:2014tc} have started simulations and BMWc achieved the first
complete simulation program~\citep{Borsanyi:2014jba}. The specificities of each
of the projects mentioned in this section is summarized in \tabref{tab:collab}.
\begin{table}
    \begin{center}
    \begin{tabular}{|r|c|c|c|c|}
        \hline
        collaboration       & RBC-UKQCD & PACS-CS & QCDSF-UKQCD & BMWc \\
        references &~\citep{Ishikawa:2012iw} &~\citep{Aoki:2012uu}
        &~\citep{Horsley:2013ti,Schierholz} &~\citep{Borsanyi:2014jba} \\
        \hline
        fermion action      & domain wall & Wilson & Wilson & Wilson\\
        $N_f$               & $2+1$     & $1+1+1$  & $1+1+1$  & $1+1+1+1$\\
        method              & re-weighting & re-weighting & HMC & HMC\\
        $\min(M_{\pi})~(\mega\eV)$  & $420$ & $135$ & $250$ & $195$\\
        $a~(\fm)$           & $0.11$ & $0.09$ & $0.08$ 
        & $0.06\,\text{--}\, 0.10$\\
        $N_{a}$             & $1$       & $1$    & $1$   & $4$  \\
        $L~(\fm)$    & $1.8$ & $2.9$ &  
            $1.9\,\text{--}\, 2.6$ & $2.1\,\text{--}\, 8.3$\\
        $N_{\mathrm{vol.}}$ & $1$       & $1$    & $2$   & $11$ \\
        \hline
    \end{tabular}
    \end{center}
    \caption{Summary of full lattice QCD+QED simulation programs. The MILC
    program~\citep{Zhou:2014tc} is too preliminary to know its
    specifications and was not included in this table. The first line is the
    fermion action used. The second line is the number of flavors used in the
    gauge configuration generation. The third line gives the simulation method
    used. The forth line indicates the minimum pion mass reached. The fifth
    line is the range of lattice spacing used and the sixth line indicates
    their number. Similarly, the seventh and eighth lines are respectively the
    range and the number of lattice spatial extents used.}
    \label{tab:collab}
\end{table}
\section{Isospin corrections to the hadron spectrum}
In this section we review the different results concerning the isospin breaking
corrections to the hadron spectrum. We first discuss the ambiguity of the
separation of strong and EM contributions. Then we present the results
concerning Dashen's theorem and quark masses and finally the hadron mass
splittings.
\subsection{Separation of QCD and QED contributions}
In all present work, only the leading isospin corrections to hadron masses
are considered. These corrections can be written as follows:
\begin{equation}
  \label{eq:qedqcdsep}
  \Delta M_X=\alpha A_X+\dm B_X+\bigo(\alpha^2,\alpha\dm, \dm^2)
\end{equation}
where $\Delta M_X$ is a given isospin mass splitting and $\dm=m_u-m_d$. Then it
is tempting to simply define the leading-order QED and QCD parts of the
splitting:
\begin{equation}
  \DQEDM_X=\alpha A_X\quad\text{and}\quad\DQCDM_X=\dm B_X
\end{equation}
However, on has to be careful because of the following ambiguity: $\alpha$ and
$\dm$ depend on each other through radiative corrections. Moreover, $m_u$ and
$m_d$ individually depend on $\alpha$ with a different coefficient because of
the difference of their electric charges. Therefore, this ambiguity cannot be
directly absorbed in higher-order isospin corrections. To make properly this
QCD/QED separation, one has to provide a prescription that defines the $\dm=0$
point. The difference between two prescriptions will be
$\bigo(m_{ud}\alpha,m_{ud}\dm)$ up to higher-order isospin corrections. Thus,
with physical quark masses where $\dm\simeq m_{ud}$ this discrepancy can be
considered as higher-order isospin corrections. So as soon as a result is
produced at the physical value of $m_{ud}$ and $\dm$, it is reasonable to
consider that the separation \eqref{eq:qedqcdsep} is effectively unambiguous up
to higher-order $\bigo(1\%)$ corrections.

Several prescriptions to define the $\dm=0$ point have been proposed in previous
works. The most conceptually straightforward scheme is to renormalize the light
quark masses in a given scheme (\eg $\overline{\mathrm{MS}}$ or RI-MOM) at a
given scale and to consistently express every quantity as a function of these
renormalized masses. This prescription was used in
\citep{Blum:2007gh,Blum:2010ge,deDivitiis:2013er}. Because renormalized quark
masses can be difficult to compute, it is also interesting to consider
prescriptions based on hadron masses. In~\citep{Borsanyi:2013va}, $\dm$ was
replaced by the mass squared difference between the connected $\bar{u}u$ and
$\bar{d}d$ mesons $\DM^2=M_{\bar{u}u}^2-M_{\bar{d}d}^2$. Is it possible to show
in partially-quenched chiral perturbation theory~\citep{Bijnens:2007jk} that
for physical quark masses, $\DM^2$ is directly proportional to $\dm$ up to the
$\bigo(1\%)$ higher-order corrections. In the later work
\citep{Borsanyi:2014jba} from the same collaboration, the prescription
$\DQEDM_{\Sigma}=0$ was used, \ie the $\Sigma^+-\Sigma^-$ mass splitting is
assumed to be proportional to $\dm$. If these particles would be point-like,
one would have $\DQEDM_{\Sigma}=0$ exactly. The authors of
\citep{Borsanyi:2014jba} found $\DQEDM_{\Sigma}$ statistically consistent with
$0$ and no more than $\unit{0.2}{\mega\eV}$ with $\DM^2=0$.
\subsection{Dashen's theorem and light quark masses}
\begin{figure}[ht]
    \hspace{2cm}
    \input{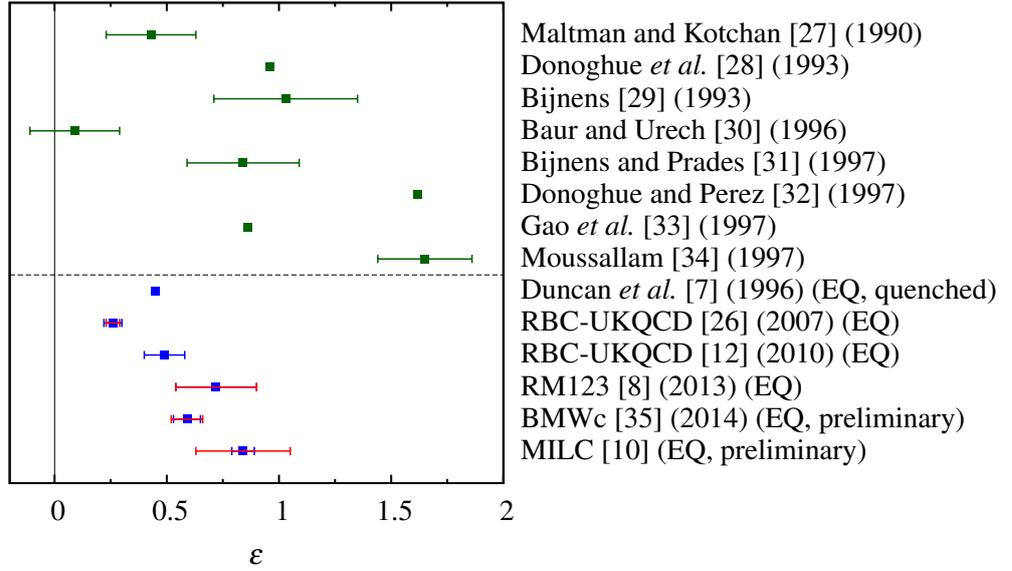}
    \vspace{2mm}
    \caption{Summary of the determination of Dashen's theorem violation
    $\epsilon$ defined in \eqref{eq:eps}. The green points represents
    analytical calculations from effective theory and the blue points results
    from lattice simulations. For the lattice results, the blue error bar is
    statistical and the red is systematic. ``EQ'' stands for electro-quenched.
    Results are presented in chronological order.}
    \label{fig:eps}
\end{figure}

Dashen's theorem corrections and individual up and down quark masses have been
computed reliably only in the electro-quenched approximation. All existing
results on Dashen's theorem correction $\epsilon$ defined in \eqref{eq:eps} are
presented in \figref{fig:eps}. Two interesting comments can be make regarding
these results. Firstly, although lattice results are still dominated by
systematic uncertainties and suffer from an uncontrolled electro-quenching
error, they look consistently spread around a common value. This contrasts
significantly with the 1990's phenomenological
determinations of this quantity. Secondly, the lattice results seems to favor a
rather large $\bigo(70\%)$ violation of Dashen's theorem.

\begin{figure}[ht]
    \hspace{2cm}
    \input{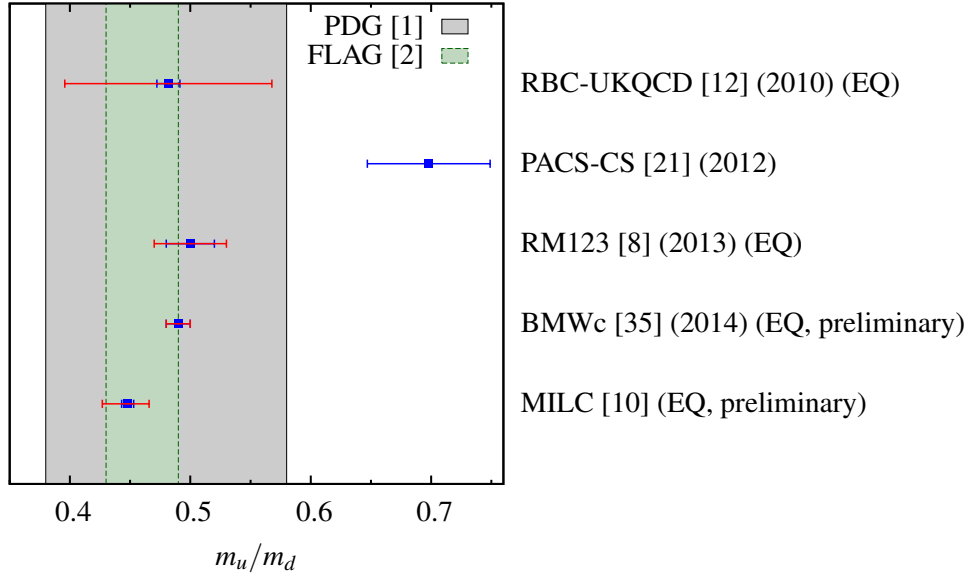}
    \vspace{2mm}
    \caption{Summary of the calculations of $m_u/m_d$ light quark mass ratio.
    Plotting conventions are identical to the one used in \figref{fig:eps}.}
    \label{fig:muomd}
\end{figure}

Regarding the light quark mass ratio $m_u/m_d$, the lattice determinations of
this number are summarized in \figref{fig:muomd}. Although there is a slight
tension between the two most recent results~\citep{Basak:2014td,BMW:2015}, they
both agree nicely with the values from the PDG~\citep{PDG2014} and
FLAG~\citep{FLAG2014} reviews. The only full QCD+QED result from
PACS-CS~\citep{Aoki:2012uu} seems to deviate significantly from other
determinations. This number is the result of an exploratory calculation using
re-weighting techniques and has unknown systematic errors. We believe that this
effect is more a systematic effect rather than an indication of a large see
quark EM contribution. It is interesting to notice that all these results
exclude strongly the $m_u=0$ solution to the strong CP problem.
\subsection{Isospin mass splittings in the hadron spectrum}
\begin{figure}[ht]
  \centering
    \includegraphics[width=0.7\linewidth]{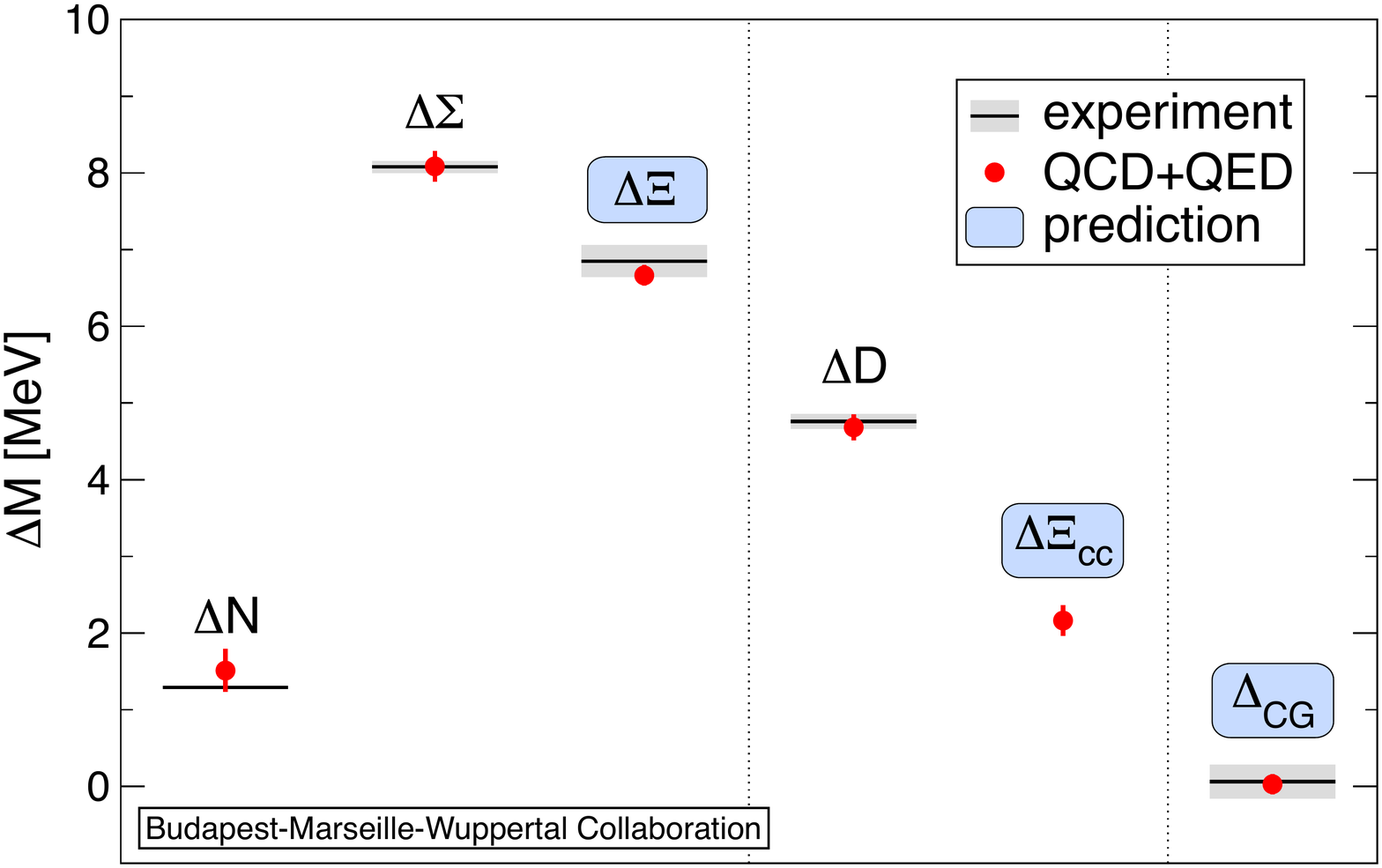}
  \caption{Summary of the results from~\citep{Borsanyi:2014jba}.
  $\Delta_{\mathrm{CG}}=\DM_N-\DM_{\Sigma}+\DM_{\Xi}$ is the correction to the
  Coleman-Glashow relation.}
  \label{fig:bmwsum}
\end{figure}

The main novelty concerning the calculation of the isospin correction to the
hadron spectrum is the high-precision determination of isospin mass splittings
of the octet baryons, the $D$ meson and the newly discovered $\Xi_{cc}$, from
the BMWc group~\citep{Borsanyi:2014jba}. These splittings where computed using
full QCD+QED simulations including an active charm quark in the sea. A summary
of these results can be found in \figref{fig:bmwsum}. The splittings obtained
in this work are in very good agreement with experimental values. It is
interesting to notice that the $\Xi$ baryon splitting is obtained with a
precision higher than the experimental measurement. Moreover, the unknown
$\Xi_{cc}$ splitting needed by charm spectrum
experiments\footnote{\cf~\eg~\href{http://www.ectstar.eu/sites/www.ectstar.eu/fi
 les/talks/after-jurgen-feb13.pdf}
{http://www.ectstar.eu/sites/www.ectstar.eu/files/talks/after-jurgen-feb13.pdf}
} is predicted accurately.

The QCDSF-UKQCD collaboration also aim at studying isospin corrections to the
octet baryon spectrum. This group have started generating full QCD+QED gauge
ensembles~\citep{Horsley:2013ti,Schierholz} to determine the corrections to the
spectrum. The analysis is performed using the same technique based on $\SU(3)$
flavor symmetry as used in their previous pure QCD work~\citep{Horsley:2012ue}.
The same collaboration also achieved the first lattice determination of the
mixing between $\Sigma^0$ and $\Lambda^0$ baryons~\citep{Horsley:2014vq} which
is authorized once isospin symmetry is broken. They obtained the mass splitting
between the two particles with a precision of $\bigo(10\%)$, in good agreement
with the experimental value.

\begin{figure}[ht]
    \centering
        \input{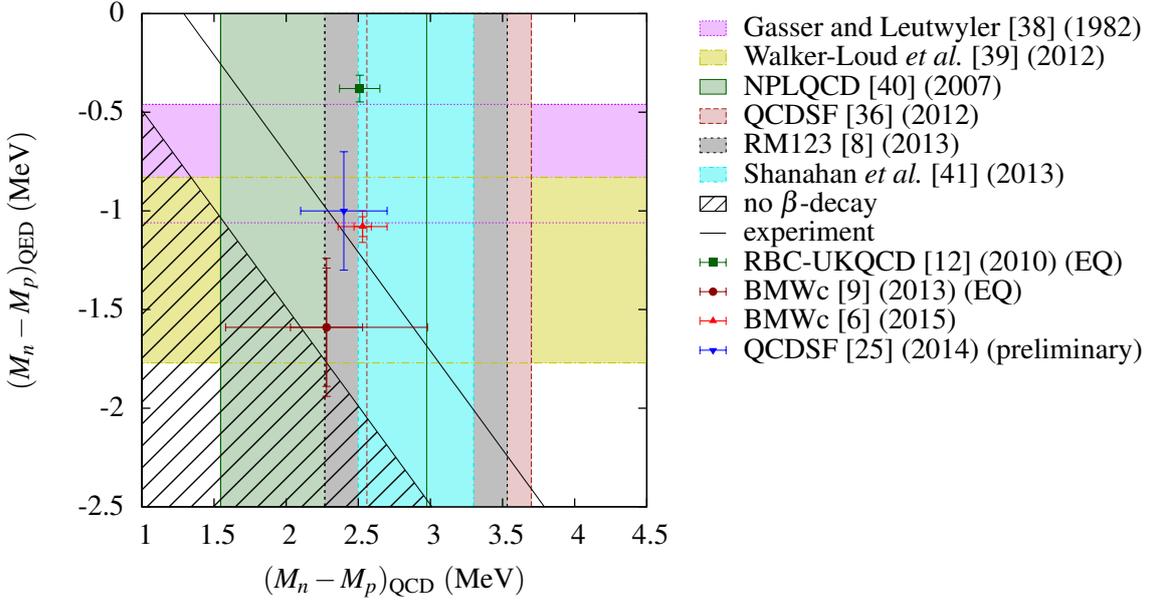}
    \caption{Review of the theoretical determinations of the nucleon mass
    splitting. Results represented by a band correspond to works where only the
    QCD or QED contribution has been determined. Where possible, error bars are
    represent statistical and systematic uncertainties. The ``no
    $\beta$-decay'' region is defined by $M_n-M_p<m_e$ where $m_e$ is the
    electron mass. ``EQ'' stands for electro-quenched.}
    \label{fig:compN}
\end{figure}

Finally we summarize the theoretical determination of the nucleon splitting,
including the QCD and QED separation, in \figref{fig:compN}. The most recent
full QCD+QED results~\citep{Borsanyi:2014jba,Schierholz} are in very good
agreement and indicate that the crucial value of the nucleon mass splitting is
indeed the result of a subtle cancellation between the QCD and QED
contributions.
\section{Conclusion and perspective}
Lattice QCD simulations in the isospin limit are reaching a precision of
$\bigo(1\%)$ and below on important observables and isospin breaking effects
are becoming the dominant source of systematic uncertainty. Therefore, it is
becoming crucial to introduce isospin breaking effects in lattice simulation in
order to provide more stringent theoretical constraints on the Standard Model.
The main challenge in this task is the inclusion of EM interactions.

The difficulty with QED comes from the IR singular structure of the theory.
Defining correctly the theory in a box is non-trivial for several reasons.
Firstly, momentum quantization can introduce hard singularities coming from the
photon field zero-mode if one uses periodic boundary conditions. Subtracting
the zero-mode of the field is a possible solution. However, fixing the
zero-mode constitutes a non-local constraint on the theory which can break some
important properties such as reflection positivity. It is shown
in~\citep{Borsanyi:2014jba} that the zero-mode subtraction proposed by Hayakawa
\& Uno~\citep{Hayakawa:2008ci} has a correct quantum mechanical interpretation.
Beyond the zero-mode subtraction, QED in a finite volume suffers from large,
power-like finite-size effects because of the long-range of the interaction.
These effects are now well understood for hadron masses. The two first orders
in the infinite-volume expansion are entirely determined by gauge invariance
and can be computed analytically~\citep{Borsanyi:2014jba,Davoudi:2014kp}.
Several effective theory descriptions of the higher-order, structure dependent
corrections have been worked
out~\citep{Hayakawa:2008ci,Blum:2010ge,Davoudi:2014kp,Fodor:2015wk}.

In its recent work, the BMWc group~\citep{Borsanyi:2014jba} achieved the first
complete lattice calculation featuring full QCD and QED interactions, an active
sea charm quark and a full control over the different sources of uncertainty.
The corrections to the baryon octet and charm spectrum are obtained precisely,
in good agreement with experimental measurements. These simulations represent
an important step toward fully non-perturbative, high precision, predictions of
Standard Model observables. The corrections to the light meson
spectrum, necessary to determine the individual up and down light quark masses
are still known only through the electro-quenched approximation.

It is now important to consider more complex quantities than hadron masses.
Adding isospin breaking effects to the determination of hadronic decay widths
is crucial to obtain high precision constraints on the flavor structure of the
Standard Model (\eg through sub-percent determinations of the CKM matrix
coefficients). With EM interactions, matrix elements are significantly harder
to determine than energy levels. Indeed, such quantity can feature IR
divergences that are physically cancelled by the addition of real soft photons
in the final state. Recently, a proposal has been made~\citep{Carrasco:2015uo}
to deal with such divergences in the case of meson decays. Also, QCD+QED
simulations can be used to perform non-perturbative computation of the hadronic
corrections to the muon anomalous magnetic moment~\citep{Blum:2015co}. This
quantity features an interesting discrepancy between theory and experiment and
it is important to reinforce the theoretical prediction to support the
experimental effort.
\acknowledgments{I am supported by the UK STFC grants ST/J000396/1 and
ST/L000296/1. I would like to thank the members of the
Budapest-Marseille-Wuppertal collaboration for their precious help and support.
I also thank C.~Bernard, R.~Horsley, G.~Schierholz and P.~Shanahan for kindly
sharing their results with me at the time of the conference. I am grateful to
L.~Del Debbio, C.~Lehner, T.~Izubuchi, A.~Patella, A.~Ramos, C.T.~Sachrajda and F.~Sanfilippo
for useful discussions.}
\bibliographystyle{pos}
\bibliography{article}

\begin{thebibliography}{42}
\providecommand{\natexlab}[1]{#1}
\providecommand{\url}[1]{\texttt{#1}}
\providecommand{\urlprefix}{}
\providecommand{\eprint}[2][]{\url{#2}}

\bibitem[{Olive \emph{et~al.}(2014)}]{PDG2014}
K.~Olive \emph{et~al.} (PDG).
\newblock \emph{{Review of Particle Physics}}.
\newblock \emph{Chin. Phys. C} \textbf{38} p. 090001 (2014).
\newblock
  \mbox{\href{http://dx.doi.org/10.1088/1674-1137/38/9/090001}{\texttt{doi:10.1088/1674-1137/38/9/090001}}}

\bibitem[{Aoki \emph{et~al.}(2014)}]{FLAG2014}
S.~Aoki \emph{et~al.} (FLAG).
\newblock \emph{{Review of lattice results concerning low-energy particle
  physics}}.
\newblock \emph{Eur. Phys. J. C} \textbf{74}(9) p. 2890 (2014).
\newblock
  \mbox{\href{http://dx.doi.org/10.1140/epjc/s10052-014-2890-7}{\texttt{doi:10.1140/epjc/s10052-014-2890-7}}}

\bibitem[{Dashen(1969)}]{Dashen:1969tn}
R.~Dashen.
\newblock \emph{{Chiral $\SU(3)\otimes\SU(3)$ as a symmetry of the strong
  interactions}}.
\newblock \emph{Phys. Rev.} \textbf{183}(5) pp. 1245--1260 (1969).
\newblock
  \mbox{\href{http://dx.doi.org/10.1103/PhysRev.183.1245}{\texttt{doi:10.1103/PhysRev.183.1245}}}

\bibitem[{Portelli \emph{et~al.}(2010)}]{Portelli:2010tz}
A.~Portelli \emph{et~al.} (BMWc).
\newblock \emph{{Electromagnetic corrections to light hadron masses}}.
\newblock \emph{PoS(Lattice 2010)121} (2010).
\newblock
  \mbox{\href{http://arxiv.org/abs/1011.4189}{\texttt{arXiv:1011.4189}}}

\bibitem[{Davoudi and Savage(2014)}]{Davoudi:2014kp}
Z.~Davoudi and M.~J. Savage.
\newblock \emph{{Finite-volume electromagnetic corrections to the masses of
  mesons, baryons, and nuclei}}.
\newblock \emph{Phys. Rev. D} \textbf{90}(5) p. 054503 (2014).
\newblock
  \mbox{\href{http://dx.doi.org/10.1103/PhysRevD.90.054503}{\texttt{doi:10.1103/PhysRevD.90.054503}}}

\bibitem[{Borsanyi \emph{et~al.}(2015)}]{Borsanyi:2014jba}
S.~Borsanyi \emph{et~al.} (BMWc).
\newblock \emph{{Ab initio calculation of the neutron-proton mass difference}}.
\newblock \emph{Science} \textbf{347} (2015).
\newblock
  \mbox{\href{http://dx.doi.org/10.1126/science.1257050}{\texttt{doi:10.1126/science.1257050}}}

\bibitem[{Duncan \emph{et~al.}(1996)Duncan, Eichten and
  Thacker}]{Duncan:1996cy}
A.~Duncan, E.~Eichten and H.~Thacker.
\newblock \emph{{Electromagnetic splittings and light quark masses in lattice
  QCD}}.
\newblock \emph{Phys. Rev. Lett.} \textbf{76}(21) pp. 3894--3897 (1996).
\newblock
  \mbox{\href{http://dx.doi.org/10.1103/PhysRevLett.76.3894}{\texttt{doi:10.1103/PhysRevLett.76.3894}}}

\bibitem[{de~Divitiis \emph{et~al.}(2013)}]{deDivitiis:2013er}
G.~M. de~Divitiis \emph{et~al.} (RM123).
\newblock \emph{{Leading isospin breaking effects on the lattice}}.
\newblock \emph{Phys. Rev. D} \textbf{87}(1) p. 114505 (2013).
\newblock
  \mbox{\href{http://dx.doi.org/10.1103/PhysRevD.87.114505}{\texttt{doi:10.1103/PhysRevD.87.114505}}}

\bibitem[{Borsanyi \emph{et~al.}(2013)}]{Borsanyi:2013va}
S.~Borsanyi \emph{et~al.} (BMWc).
\newblock \emph{{Isospin splittings in the light baryon octet from lattice QCD
  and QED}}.
\newblock \emph{Phys. Rev. Lett.} \textbf{111}(25) p. 252001 (2013).
\newblock
  \mbox{\href{http://dx.doi.org/10.1103/PhysRevLett.111.252001}{\texttt{doi:10.1103/PhysRevLett.111.252001}}}

\bibitem[{Basak \emph{et~al.}(2014)}]{Basak:2014td}
S.~Basak \emph{et~al.} (MILC).
\newblock \emph{{Finite-volume effects and the electromagnetic contributions to
  kaon and pion masses}}.
\newblock \emph{PoS(Lattice 2014)} (2014).
\newblock
  \mbox{\href{http://arxiv.org/abs/1409.7139}{\texttt{arXiv:1409.7139}}}

\bibitem[{Hayakawa and Uno(2008)}]{Hayakawa:2008ci}
M.~Hayakawa and S.~Uno.
\newblock \emph{{QED in finite volume and finite size scaling effect on
  electromagnetic properties of hadrons}}.
\newblock \emph{Prog. Theor. Phys.} \textbf{120}(3) pp. 413--441 (2008).
\newblock
  \mbox{\href{http://dx.doi.org/10.1143/PTP.120.413}{\texttt{doi:10.1143/PTP.120.413}}}

\bibitem[{Blum \emph{et~al.}(2010)}]{Blum:2010ge}
T.~Blum \emph{et~al.} (RBC-UKQCD).
\newblock \emph{{Electromagnetic mass splittings of the low lying hadrons and
  quark masses from 2+1 flavor lattice QCD+QED}}.
\newblock \emph{Phys. Rev. D} \textbf{82} p. 094508 (2010).
\newblock
  \mbox{\href{http://dx.doi.org/10.1103/PhysRevD.82.094508}{\texttt{doi:10.1103/PhysRevD.82.094508}}}

\bibitem[{Fodor \emph{et~al.}(2015)}]{Fodor:2015wk}
Z.~Fodor \emph{et~al.}
\newblock \emph{{Quantum electrodynamics in finite volume and nonrelativistic
  effective field theories}}  (2015).
\newblock
  \mbox{\href{http://arxiv.org/abs/1502.06921}{\texttt{arXiv:1502.06921}}}

\bibitem[{Zhou and Gottlieb(2014)}]{Zhou:2014tc}
R.~Zhou and S.~Gottlieb (MILC).
\newblock \emph{{Dynamical QCD+QED simulation with staggered quarks}}.
\newblock \emph{PoS(Lattice 2014)} (2014).
\newblock
  \mbox{\href{http://arxiv.org/abs/1411.4115}{\texttt{arXiv:1411.4115}}}

\bibitem[{Seiler \emph{et~al.}(1984)Seiler, Stamatescu and
  Zwanziger}]{Seiler:1984ju}
E.~Seiler, I.~O. Stamatescu and D.~Zwanziger.
\newblock \emph{{Monte Carlo simulation of non-compact QCD with stochastic
  gauge fixing}}.
\newblock \emph{Nucl. Phys. B} \textbf{239}(1) pp. 177--200 (1984).
\newblock
  \mbox{\href{http://dx.doi.org/10.1016/0550-3213(84)90089-0}{\texttt{doi:10.1016/0550-3213(84)90089-0}}}

\bibitem[{Bijnens and Danielsson(2007)}]{Bijnens:2007jk}
J.~Bijnens and N.~Danielsson.
\newblock \emph{{Electromagnetic corrections in partially quenched chiral
  perturbation theory}}.
\newblock \emph{Phys. Rev. D} \textbf{75}(1) p. 014505 (2007).
\newblock
  \mbox{\href{http://dx.doi.org/10.1103/PhysRevD.75.014505}{\texttt{doi:10.1103/PhysRevD.75.014505}}}

\bibitem[{Portelli \emph{et~al.}(2012)}]{Portelli:2012vz}
A.~Portelli \emph{et~al.} (BMWc).
\newblock \emph{{Systematic errors in partially-quenched QCD plus QED lattice
  simulations}}.
\newblock \emph{PoS(Lattice 2011)136} (2012).
\newblock
  \mbox{\href{http://arxiv.org/abs/1201.2787}{\texttt{arXiv:1201.2787}}}

\bibitem[{Portelli(2013)}]{Portelli:2013wf}
A.~Portelli.
\newblock \emph{{Review on the inclusion of isospin breaking effects in lattice
  calculations}}.
\newblock \emph{PoS(KAON13)023} (2013).
\newblock
  \mbox{\href{http://arxiv.org/abs/1307.6056}{\texttt{arXiv:1307.6056}}}

\bibitem[{Duncan \emph{et~al.}(2005)Duncan, Eichten and
  Sedgewick}]{Duncan:2005cq}
A.~Duncan, E.~Eichten and R.~Sedgewick.
\newblock \emph{{Computing Electromagnetic Effects in Fully Unquenched QCD}}
  \textbf{71}(9) p. 094509 (2005).
\newblock
  \mbox{\href{http://dx.doi.org/10.1103/PhysRevD.71.094509}{\texttt{doi:10.1103/PhysRevD.71.094509}}}

\bibitem[{Ishikawa \emph{et~al.}(2012)}]{Ishikawa:2012iw}
T.~Ishikawa \emph{et~al.}
\newblock \emph{{Full QED+QCD low-energy constants through reweighting}}.
\newblock \emph{Physical Review Letters} \textbf{109}(7) p. 072002 (2012).
\newblock
  \mbox{\href{http://dx.doi.org/10.1103/PhysRevLett.109.072002}{\texttt{doi:10.1103/PhysRevLett.109.072002}}}

\bibitem[{Aoki \emph{et~al.}(2012)}]{Aoki:2012uu}
S.~Aoki \emph{et~al.} (PACS-CS).
\newblock \emph{{1+1+1 flavor QCD + QED simulation at the physical point}}
  \textbf{86}(3) p. 034507 (2012).
\newblock
  \mbox{\href{http://dx.doi.org/10.1103/PhysRevD.86.034507}{\texttt{doi:10.1103/PhysRevD.86.034507}}}

\bibitem[{Francis \emph{et~al.}(2014)}]{Francis:2014th}
A.~Francis \emph{et~al.}
\newblock \emph{{The leading disconnected contribution to the anomalous
  magnetic moment of the muon}}.
\newblock \emph{Proceedings of the XXXII International Symposium on Lattice
  Field Theory, PoS(Lattice 2014)} (2014).
\newblock
  \mbox{\href{http://arxiv.org/abs/1411.7592}{\texttt{arXiv:1411.7592}}}

\bibitem[{Carrasco \emph{et~al.}(2015)}]{Carrasco:2015uo}
N.~Carrasco \emph{et~al.}
\newblock \emph{{QED Corrections to Hadronic Processes in Lattice QCD}}
  (2015).
\newblock
  \mbox{\href{http://arxiv.org/abs/1502.00257}{\texttt{arXiv:1502.00257}}}

\bibitem[{Horsley \emph{et~al.}(2013)}]{Horsley:2013ti}
R.~Horsley \emph{et~al.} (QCDSF-UKQCD).
\newblock \emph{{Electromagnetic splitting of quark and pseudoscalar meson
  masses from dynamical QCD + QED}}.
\newblock \emph{PoS(LATTICE 2013)499} (2013).
\newblock
  \mbox{\href{http://arxiv.org/abs/1311.4554}{\texttt{arXiv:1311.4554}}}

\bibitem[{Schierholz(2014)}]{Schierholz}
G.~Schierholz (QCDSF-UKQCD).
\newblock \emph{{Private communication}}  (2014)

\bibitem[{Blum \emph{et~al.}(2007)}]{Blum:2007gh}
T.~Blum \emph{et~al.} (RBC-UKQCD).
\newblock \emph{{Determination of light quark masses from the electromagnetic
  splitting of pseudoscalar meson masses computed with two flavors of domain
  wall fermions}}.
\newblock \emph{Phys. Rev. D} \textbf{76}(11) p. 114508 (2007).
\newblock
  \mbox{\href{http://dx.doi.org/10.1103/PhysRevD.76.114508}{\texttt{doi:10.1103/PhysRevD.76.114508}}}

\bibitem[{Maltman and Kotchan(1990)}]{Maltman:1990dh}
K.~Maltman and D.~Kotchan.
\newblock \emph{{Chiral log corrections to pseudoscalar electromagnetic
  self-energies and the reliability of Dashen's theorem}}.
\newblock \emph{Mod. Phys. Lett. A} \textbf{5}(29) pp. 2457--2464 (1990).
\newblock
  \mbox{\href{http://dx.doi.org/10.1142/S0217732390002857}{\texttt{doi:10.1142/S0217732390002857}}}

\bibitem[{Donoghue \emph{et~al.}(1993)Donoghue, Holstein and
  Wyler}]{Donoghue:1993bm}
J.~F. Donoghue, B.~R. Holstein and D.~Wyler.
\newblock \emph{{Electromagnetic self-energies of pseudoscalar mesons and
  Dashen's theorem}}.
\newblock \emph{Phys. Rev. D} \textbf{47}(5) pp. 2089--2097 (1993).
\newblock
  \mbox{\href{http://dx.doi.org/10.1103/PhysRevD.47.2089}{\texttt{doi:10.1103/PhysRevD.47.2089}}}

\bibitem[{Bijnens(1993)}]{Bijnens:1993go}
J.~Bijnens.
\newblock \emph{{Violations of Dashen's theorem}}.
\newblock \emph{Phys. Lett. B} \textbf{306}(3-4) pp. 343--349 (1993).
\newblock
  \mbox{\href{http://dx.doi.org/10.1016/0370-2693(93)90089-Z}{\texttt{doi:10.1016/0370-2693(93)90089-Z}}}

\bibitem[{Baur and Urech(1996)}]{Baur:1996gf}
R.~Baur and R.~Urech.
\newblock \emph{{Corrections to Dashen's theorem}}.
\newblock \emph{Phys. Rev. D} \textbf{53}(11) pp. 6552--6557 (1996).
\newblock
  \mbox{\href{http://dx.doi.org/10.1103/PhysRevD.53.6552}{\texttt{doi:10.1103/PhysRevD.53.6552}}}

\bibitem[{Bijnens and Prades(1997)}]{Bijnens:1997ku}
J.~Bijnens and J.~Prades.
\newblock \emph{{Electromagnetic corrections for pions and kaons : masses and
  polarizabilities}}.
\newblock \emph{Nucl. Phys. B} \textbf{490}(1-2) pp. 239--271 (1997).
\newblock
  \mbox{\href{http://dx.doi.org/10.1016/S0550-3213(97)00107-7}{\texttt{doi:10.1016/S0550-3213(97)00107-7}}}

\bibitem[{Donoghue and Perez(1997)}]{Donoghue:1997kt}
J.~F. Donoghue and A.~F. Perez.
\newblock \emph{{The electromagnetic mass differences of pions and kaons}}.
\newblock \emph{Phys. Rev. D} \textbf{55}(11) pp. 7075--7092 (1997).
\newblock
  \mbox{\href{http://dx.doi.org/10.1103/PhysRevD.55.7075}{\texttt{doi:10.1103/PhysRevD.55.7075}}}

\bibitem[{Gao \emph{et~al.}(1997)Gao, Li and Yan}]{Gao:1997hq}
D.-N. Gao, B.~A. Li and M.-L. Yan.
\newblock \emph{{Electromagnetic mass splittings of $\pi$, $a_{1}$, $K$,
  $K_{1}(1400)$, and $K^{*}(892)$}}.
\newblock \emph{Phys. Rev. D} \textbf{56}(7) pp. 4115--4132 (1997).
\newblock
  \mbox{\href{http://dx.doi.org/10.1103/PhysRevD.56.4115}{\texttt{doi:10.1103/PhysRevD.56.4115}}}

\bibitem[{Moussallam(1997)}]{Moussallam:1997jm}
B.~Moussallam.
\newblock \emph{{A Sum rule approach to the violation of Dashen's theorem}}.
\newblock \emph{Nucl. Phys. B} \textbf{504}(1-2) pp. 381--414 (1997).
\newblock
  \mbox{\href{http://dx.doi.org/10.1016/S0550-3213(97)00464-1}{\texttt{doi:10.1016/S0550-3213(97)00464-1}}}

\bibitem[{{Budapest-Marseille-Wuppertal collaboration}(in
  preparation)}]{BMW:2015}
{Budapest-Marseille-Wuppertal collaboration}.
\newblock \emph{{Up and down quark masses and corrections to Dashen’s theorem
  from lattice QCD and QED}}  (in preparation)

\bibitem[{Horsley \emph{et~al.}(2012)}]{Horsley:2012ue}
R.~Horsley \emph{et~al.} (QCDSF-UKQCD).
\newblock \emph{{Isospin breaking in octet baryon mass splittings}}.
\newblock \emph{Phys. Rev. D} \textbf{86}(44) p. 114511 (2012).
\newblock
  \mbox{\href{http://dx.doi.org/10.1103/PhysRevD.86.114511}{\texttt{doi:10.1103/PhysRevD.86.114511}}}

\bibitem[{Horsley \emph{et~al.}(2014)}]{Horsley:2014vq}
R.~Horsley \emph{et~al.}
\newblock \emph{{A lattice determination of Sigma - Lambda mixing}}  (2014).
\newblock
  \mbox{\href{http://arxiv.org/abs/1411.7665}{\texttt{arXiv:1411.7665}}}

\bibitem[{Gasser and Leutwyler(1982)}]{Gasser:1982ic}
J.~Gasser and H.~Leutwyler.
\newblock \emph{{Quark masses}}.
\newblock \emph{Phys. Rep.} \textbf{87}(3) pp. 77--169 (1982).
\newblock
  \mbox{\href{http://dx.doi.org/10.1016/0370-1573(82)90035-7}{\texttt{doi:10.1016/0370-1573(82)90035-7}}}

\bibitem[{Walker-Loud \emph{et~al.}(2012)Walker-Loud, Carlson and
  Miller}]{WalkerLoud:2012gq}
A.~Walker-Loud, C.~E. Carlson and G.~A. Miller.
\newblock \emph{{The electromagnetic self-energy contribution to $M_p-M_n$ and
  the isovector nucleon magnetic polarizability}}.
\newblock \emph{Phys. Rev. Lett.} \textbf{108}(23) p. 232301 (2012).
\newblock
  \mbox{\href{http://dx.doi.org/10.1103/PhysRevLett.108.232301}{\texttt{doi:10.1103/PhysRevLett.108.232301}}}

\bibitem[{Beane \emph{et~al.}(2007)Beane, Orginos and Savage}]{Beane:2007eh}
S.~R. Beane, K.~Orginos and M.~J. Savage (NPLQCD).
\newblock \emph{{Strong-isospin violation in the neutron-proton mass difference
  from fully-dynamical lattice QCD and PQQCD}}.
\newblock \emph{Nucl. Phys. B} \textbf{768}(1-2) pp. 38--50 (2007).
\newblock
  \mbox{\href{http://dx.doi.org/10.1016/j.nuclphysb.2006.12.023}{\texttt{doi:10.1016/j.nuclphysb.2006.12.023}}}

\bibitem[{Shanahan \emph{et~al.}(2013)Shanahan, Thomas and
  Young}]{Shanahan:2013kg}
P.~E. Shanahan, A.~W. Thomas and R.~D. Young.
\newblock \emph{{Strong contribution to octet baryon mass splittings}}.
\newblock \emph{Phys. Lett. B} \textbf{718}(3) pp. 1148--1153 (2013).
\newblock
  \mbox{\href{http://dx.doi.org/10.1016/j.physletb.2012.11.072}{\texttt{doi:10.1016/j.physletb.2012.11.072}}}

\bibitem[{Blum \emph{et~al.}(2015)Blum, Chowdhury, Hayakawa and
  Izubuchi}]{Blum:2015co}
T.~Blum, S.~Chowdhury, M.~Hayakawa and T.~Izubuchi.
\newblock \emph{{Hadronic Light-by-Light Scattering Contribution to the Muon
  Anomalous Magnetic Moment from Lattice QCD}}.
\newblock \emph{Phys. Rev. Lett.} \textbf{114}(1) p. 012001 (2015).
\newblock
  \mbox{\href{http://dx.doi.org/10.1103/PhysRevLett.114.012001}{\texttt{doi:10.1103/PhysRevLett.114.012001}}}

\end{thebibliography}

\end{document}